\documentclass[12pt, letterpaper, ]{article}

\usepackage[utf8]{inputenc}
\usepackage{amssymb}
\usepackage{multicol}
\usepackage[margin=2.35cm]{geometry}
\usepackage{setspace}
\usepackage{url}
\usepackage{color}
\usepackage{tikz}
\usepackage{amsmath}
\usepackage{physics}

\begin{document}

\title{Carnap on Quantum Mechanics}

\author{Sebastian Horvat\thanks{Department of Philosophy, University of Vienna} \and Iulian D. Toader\thanks{Department of Philosophy \& Institute Vienna Circle, University of Vienna \newline Invited contribution forthcoming in \textit{Carnap Handbuch}, Metzler, ed. by Georg Schiemer and Christian Damböck.}}

\date{}

\maketitle

\onehalfspacing

\bigskip

Central to Carnap’s view on quantum mechanics (QM), as expressed in his 1966 book, \textit{Philosophical Foundations of Physics}, is the somewhat cautious claim that philosophical questions about the nature and implications of QM, — especially questions about its language and logic — cannot properly be addressed until the theory has been presented as a formalized axiomatic system, which he took to be still a work-in-progress of the scientific community. This might suggest that, unlike his reflections on other theories, Carnap’s thoughts on QM were not entirely up to date, which is what the editors of a recent anthology emphasize (Lutz and Tuboly, 2021, 5). To be sure, Carnap seems to have implied that, for example, John von Neumann’s treatment of the theory, in his 1932 book, \textit{Mathematical Foundations of Quantum Mechanics}, is insufficient as a basis for answering philosophical questions of the kind Carnap thought should be asked about QM. Furthermore, although his 1966 book is based on lectures given in the 1950s, he did not take into account other rigorous approaches, like George W. Mackey’s 1963 book, \textit{Mathematical Foundations of Quantum Mechanics}, which would arguably have satisfied, to a larger extent, Carnap’s conditions for proper philosophical analysis. Indeed, he also appears to have ignored what would become foundational milestones of QM (e.g., the theorems proved in Gleason, 1957 and Bell, 1964). That Carnap, nevertheless, had a sound understanding of the basic principles of QM is beyond doubt, and claims to the effect that he was only “familiar with relativity theory” and not with quantum theory (Faye and Jaksland, 2021, 118) are simply false.

We will start by briefly describing Carnap’s own presentation of these principles, and the answers that he thought they suggested to some philosophical questions like the nature of scientific explanation and fundamental ontology, the meaning of theoretical terms, and the existence of free will. Then we shall turn to the issues that, on Carnap’s view,  could not yet be properly addressed by philosophers, like the nature of the language and logic of QM, and to the metaphilosophy that made him believe that they could not be so addressed. We will then end by drawing on his work on the semantics of classical logic and on the epistemology of rational reconstruction in order to formulate some questions that Carnap would have probably addressed concerning some contemporary approaches to QM.

“Indeterminism in Quantum Physics” — the very last chapter of (Carnap 1966) — starts by presenting some basic principles of QM. Heisenberg’s Uncertainty Principle is characterized as “a fundamental law that must hold as long as the laws of quantum theory are maintained in their present form” (op. cit., 284). Carnap duly noted that the limitations entailed by this principle cannot be reduced through any possible improvements of our measuring techniques, since they are not due to the imperfections of our measuring instruments. The mathematical representation of a quantum state by means of a wave function defined on an abstract higher-dimensional space, i.e. on configuration space, is carefully presented. Carnap described the deterministic dynamics of quantum-mechanical systems, governed by the Schrödinger equation, and the probabilistic character of all predictions of the results of any measurements performed on such systems, briefly touching on the QM of macroscopic objects (like satellites) as well. On the basis of his understanding of the basic principles of QM, Carnap offered, throughout the book, his answers to some important philosophical questions, which he obviously thought could already be addressed on that basis.

Thus, in Carnap’s view, the necessarily statistical character of quantum-mechanical explanations cannot be regarded as a manifestation of our ignorance, but as an expression of the basic structure of the world, which entails that all physical explanations can only be statistical, under the assumption that all laws of physics reduce to the fundamental principles of QM (such as the Uncertainty Principle) (Carnap 1966, 9). Relatedly, Carnap reported as an “interesting speculation” that QM might indicate that this very structure, and thus presumably its fundamental ontology, including space and time, is all discrete, rather than continuous (Carnap 1966, 89). Furthermore, according to Carnap, QM clearly suggests that there can be no explicit definitions of quantum-theoretical concepts in terms of empirical concepts. But he noted that a more satisfactory answer to the question about the empirical meaning of quantum properties like spin, for instance, would require “an elaborate theory” (Carnap 1966, 221), by which he meant a formalized axiomatization of QM, i.e., what he had called a physical calculus together with a set of semantic or correspondence rules that connect this calculus to its empirical interpretations (Carnap 1939, 60). Carnap also argued that QM is irrelevant to philosophical debates on the existence of free will. This is because, in his view, indeterminate quantum jumps, though random, cannot play any role in decision-making since “it is not likely that these are points at which human decisions are made” (Carnap 1966, 221). But even if they were, that would only make our decisions equally random, and so they would simply not be choices at all, but chances. And even if the range of quantum randomness were much greater than in the actual world, as described by QM, that would only decrease the possibility of free choices. This issue has been since reconsidered, of course, especially by defenders of libertarianism (see, e.g., Kane, 2014).

There are, according to Carnap, also questions that cannot be addressed properly on the basis of the formulations of QM that Carnap was aware of. These include questions about the logic and language of the theory. He wrote: “The revolutionary nature of the Heisenberg uncertainty principle has led some philosophers and physicists to suggest that certain basic changes be made in the language of physics. [...] The most extreme proposals for such modification concern a change in the form of logic used in physics.” (Carnap 1966, 288) Among these proposals, he recalled Martin Strauss’ revision of the formation rules in the language of QM, motivated by the meaninglessness of classical conjunctions of statements about conjugate observables, like momentum and position (for discussion of Carnap’s reply to Strauss and his exchanges with Bohr, see Faye and Jaksland, 2021). Carnap further mentioned Garrett Birkhoff and von Neumann’s (1936) change of the transformation rules, by the replacement of the law of distribution (of conjunction over disjunction) by that of modularity, as well as Reichenbach’s (1944) replacement of the law of the excluded third by that of the excluded fourth. However, unlike Putnam (1968), Carnap was not ready to take lessons in logic from QM. In a meeting in 1962 in Los Angeles, upon hearing Putnam’s endorsement of Birkhoff and Von Neumann’s non-distributive logic, Carnap responded in the following way: “Ich sage, dass ich eher bereit wäre, weitgehende Änderungen in den physikalischen Begriffen zu machen”.\footnote{Presumably Carnap would have preferred pilot-wave-like modifications of the quantum formalism, as he in fact had mentioned in an earlier conversation with Putnam in 1960 at Stanford: “Ich sage, ich habe neigung zu der Ansicht von Bohm, usw.” We thank Marij van Strien for drawing our attention to these passages in Carnap’s diary.} The question whether the logic of physics ought to be revised is not one that could be addressed properly without first presenting “the entire field of physics stated in a systematic form that would include formal logic.” (Carnap 1966, 290)

Even though he is not fully explicit about it, it is quite clear that Carnap, in accord with the tendencies that manifested throughout his philosophy starting with his work in the 1920s (e.g., Carnap 1928), demanded a \textit{rational reconstruction} of modern physics (for discussions on Carnap’s notion of rational reconstruction, see e.g. Demopoulos, 2007; Beaney, 2013). A rational reconstruction would require modern physics to be couched in terms of a partially interpreted syntactic structure that includes, along with the logico-mathematical axioms, theoretical sentences and correspondence rules that partially endow that structure with empirical meaning by relating theoretical sentences to observation sentences. As things stood at the time when Carnap wrote and published his 1966 book, he did not think that such a rational reconstruction of modern physics had been given: “[Its] language is still, except for its mathematical part, largely a natural language; that is, its rules are learned implicitly in practice and seldom formulated explicitly.” (Carnap 1966, 291) This tacitly implies that even von Neumann’s (1932) and Mackey’s (1963) rigorous formulations of QM — widely regarded as mathematical axiomatizations of QM \textit{par excellence} — fail to qualify as rational reconstructions of modern physics, despite their mathematical clarity and their transparent use of the axiomatic method.

We think that there are two main reasons that underlie this negative assessment. First, as mentioned above, a rational reconstruction requires not only the logico-mathematical structures of a scientific theory to be couched in a formal language, but its empirical aspects as well: a formal language should be introduced, whose terms are split into theoretical and observational ones, and the theory should be formulated as a conjunction of theoretical sentences and correspondence rules. Secondly, and perhaps more importantly, note that Carnap did not refer to the lack of systematization of \textit{QM in particular}, but of \textit{modern physics as a whole}: thus, even a complete rational reconstruction of QM alone, would still not provide Carnap with a proper basis for discussing alternative logico-linguistic frameworks, as that would not comprise gravitational and/or spacetime physics. Therefore, Carnap’s unwillingness to take lessons in logic from modern physics is caused not only by the absence of a fully formalized axiomatization of the latter, but also by its disunity, which is still manifest today in the tensions between quantum and gravitational physics. Nevertheless, although Carnap considered the applications of logical methods to modern physics to be still in their infancy, while recalling the vast success that such methods enjoyed in the foundations of mathematics, he expressed himself rather optimistically: “I am convinced that two tendencies [...] will prove equally effective in sharpening and clarifying the language of physics: the application of modern logic and set theory, and the adoption of the axiomatic method in its modern form, which presupposes a formalized language system.” (op. cit., 291)

Since Carnap’s 1966 book, not only has the range of applications of QM been significantly widened, including developments in particle physics, condensed matter theory and quantum information technologies, but there have also been important advances in our philosophical understanding of QM as well. The space of viable ways of interpreting its formalism has been extensively explored, bringing about novel theories and interpretations ranging from purportedly realist-friendly hidden variables and many-worlds theories to anti-realist ones such as QBism (e.g., Freire \textit{et al.}, 2022).As Gleason (1957), Bell (1964), Kochen and Specker (1967), and others have taught us, classical-like models of certain empirical phenomena must violate some (arguably) desirable features such as locality and non-contextuality, hence causing problems for “naive” realist construals of quantum phenomena. Even though some of these seminal results were already produced during the times when Carnap was active, their significance became recognized by the broader philosophical and scientific community only later, which partly explains why there are no references to them in his writings. But what would Carnap have thought about these results, had he been aware of them? What would he have to tell us today about QM?

Carnap’s views on realism in science, as presented in his works in the 50s and 60s (e.g., Carnap, 1950), indicate quite clearly that he would take disputes over which account of QM provides the correct description of what is “really out there” (be it particles, flashes, many worlds, etc.) as meaningless pseudo-disputes, since they ask “external questions”, i.e., questions that are not posed within a certain linguistic framework, but which concern the framework itself. Still, we think he would find the debates over current interpretations of QM as fruitful, if understood as non-metaphysical disputes over the best way of formulating the theory, where alternatives are evaluated on pragmatic criteria such as simplicity and amenability to unification with other areas of physics. Carnap would not interpret the no-go results of Bell, Kochen and Specker, and others, as signaling obstacles to extending realism to the microscopic domain. Rather, he would most likely see them as indicating metasemantic limitations to providing a representationalist semantics for the formalism of QM. Incidentally, Richard Healey has recently defended a view along such lines, which takes QM to require an inferentialist semantics (Healey, 2017). On this view, it is neither logic, nor ontology that makes QM revolutionary, but rather the metasemantic consideration that the rules of the theory — more exactly, the circumstances and consequences of their applications — determine the meaning assigned to its basic non-logical terms, e.g. quantum states and observables. How would Carnap have reacted to this proposal? 

As discussed in other entries of this \textit{Handbook}, whereas Carnap endorsed a representationalist semantics for non-logical, descriptive terms, he did endorse an inferentialist semantics for logical constants, as can be seen most clearly in his \textit{Logical Syntax}: “let any postulates and any rules of inference be chosen arbitrarily; then this choice, whatever it may be, will determine what meaning is to be assigned to the fundamental logical symbols.” (Carnap 1937, xv, emphasis removed). Famously, after considering classical logic as a formalized axiomatic system, he realized that its rules fail to univocally determine the meaning of logical terms (Carnap 1943). He proved the existence of non-normal interpretations of the classical logical calculus, i.e. interpretations which make the connectives non-truth-functional, e.g. by allowing true classical disjunctions with false disjuncts. Since such interpretations are not isomorphic to the normal truth-tables, this came to be considered as a categoricity problem for inferentialism about classical logic (Raatikainen, 2008), to which Carnap himself already attempted to provide a solution (Carnap, 1943). Relating this back to Healey, even if Carnap were indeed to follow him in taking the metasemantic restrictions suggested by the no-go theorems to provide a sufficient reason for extending inferentialism to the non-logical terms of QM, he would have probably applied to QM the same test that he applied to classical logic: namely, he would have wanted to find out whether the rules of the theory, as understood by Healey, or more exactly the physical rules (i.e., Carnap’s P-rules) of its rational reconstruction, are categorical;  if they turn out to be otherwise, he might have urged one to find a solution to this new categoricity problem.

In closing, recall that Carnap was rather cautious in drawing conclusions about logic from the physics of his time, as he deemed the latter’s axiomatizations to be still work-in-progress. Would his assessment of today’s physics be any different? Are there any axiomatizations of QM, of the type he demanded as a precondition for a proper philosophical analysis of its logic and language, and have they led, as he had expected, to the creation of new concepts that helped “rebuild the theory” in ways that further developed it (Carnap, 1966, 291)? There are certainly no complete rational reconstructions of modern physics, in Carnap’s sense (again, the unification of quantum and spacetime theory is an ongoing research program), nor of QM alone for that matter, so he would most likely still not be convinced to consider revising logic due to the peculiarities of QM. Nevertheless, there have been some recently developed axiomatizations of QM that Carnap might have been interested in, namely those axiomatizations that derive certain structural features of QM (e.g., its state space) from information-theoretic principles, i.e. principles concerning the (im)possibilities of encoding, transmitting and decoding information with physical systems (e.g., Hardy, 2001; Chiribella \textit{et al.}, 2011). Even though these axiomatizations in their present form do not come even close to a Carnapian rational reconstruction of QM, as they are not even couched in a formal language, they might still pave the way towards one.\\
\\
\\
\textbf{References}\\

{\setstretch{1.0}
\noindent Beaney, Michael: Analytic Philosophy and History of Philosophy: The Development of the Idea of Rational Reconstruction. In Erich H. Reck (ed.), \textit{The Historical Turn in Analytic Philosophy}. London: Palgrave-Macmillan, 2013, 231-260\\
\\
Bell, John S.: On the Einstein Podolsky Rosen Paradox. \textit{Physics, 1}, 1964, 195-200 \\
\\
Birkhoff, Garrett and von Neumann, John: The Logic of Quantum Mechanics. \textit{Annals of Mathematics 37}, 1936, 823-843\\
\\
Carnap, Rudolf: \textit{Der logische Aufbau der Welt}, Berlin-Schlachtensee: Weltkreis-Verlag, 1928; 2nd ed. Hamburg: Felix Meiner, 1961\\
\\
Carnap, Rudolf: \textit{Logical Syntax of Language}, London: K. Paul, Trench, Trubner $\&$ Co Ltd., 1937\\
\\
Carnap, Rudolf: \textit{Foundations of Logic and Mathematics}, Chicago: University of Chicago Press, 1939\\
\\
Carnap, Rudolf: \textit{Formalization of Logic}, Harvard University Press, 1943\\
\\
Carnap, Rudolf: \textit{Empiricism, Semantics and Ontology}. Revue Internationale de Philosophie, 4, 1950, 20-40\\
\\
Carnap, Rudolf: \textit{Philosophical Foundations of Physics: An Introduction to Philosophy of Science}. New York: Basic Books, 1966\\
\\
Chiribella, Giulio, Giacomo M. D’Ariano, and Paolo Perinotti: Informational derivation of quantum theory. \textit{Physical Review A 84}, 2011, 012311\\
\\
Demopoulos, William: Carnap on the Rational Reconstruction of Scientific Theories. In Michael Friedman and Richard Creath (Eds.) \textit{The Cambridge Companion to Carnap}. Cambridge: Cambridge University Press, 2007, 248-272\\
\\
Faye, Jan and Jaksland, Rasmus: What Bohr wanted Carnap to learn from quantum mechanics. \textit{Studies in History and Philosophy of Science 88}, 2021, 110–119\\
\\
Freire Jr, Olival, Guido Bacciagaluppi, Olivier Darrigol, Thiago Hartz, Christian Joas, Alexei Kojevnikov, and Osvaldo Pessoa Jr, eds.: \textit{The Oxford Handbook of the History of Quantum Interpretations}. Oxford University Press, 2022.\\
\\
Gleason, Andrew M.: Measures on the Closed Subspaces of a Hilbert Space. \textit{Journal of Mathematics and Mechanics 6}, 1957, 885-893\\
\\
Hardy, Lucien: Quantum theory from five reasonable axioms. arXiv preprint quant-ph/0101012 2001\\
\\
Healey, Richard: \textit{The Quantum Revolution in Philosophy}. Oxford: Oxford University Press, 2017\\
\\
Kane, Robert: Quantum Physics, Action and Free Will: How Might Free Will Be Possible in a Quantum Universe? In Uwe Meixner and Antonella Corradini (Eds.) \textit{Quantum Physics Meets the Philosophy of Mind: New Essays on the Mind-Body Relation in Quantum-Theoretical Perspective}. De Gruyter, 2014, 163-182\\
\\
Kochen Simon and Ernst P. Specker: The Problem of Hidden Variables in Quantum Mechanics. \textit{Journal of Mathematics and Mechanics 17}, 59-87\\
\\
Lutz, Sebastian and Tuboly, Adam Tamas: Introduction: From Philosophy of Nature to Philosophy of Physics. In Sebastian Lutz and Adam Tamas Tuboly (Eds.) \textit{Logical Empiricism and the Physical Sciences: From Philosophy of Nature to Philosophy of Physics}. New York and London: Routledge, 2021, 1-17\\
\\
Mackie, George. W.: \textit{Mathematical Foundations of Quantum Mechanics}. New York: W. A. Benjamin, 1963\\
\\
Putnam, Hilary: Is logic empirical? In R. S. Cohen and M. W. Wartofsky (Eds.) \textit{Boston Studies in the Philosophy of Science}. Dordrecht: Reidel, 1968, 216-241 \\
\\
Raatikainen, Panu: On Rules of Inference and the Meanings of Logical Constants. \textit{Analysis 68}, 2008, 282-287\\
\\
Reichenbach, Hans: \textit{Philosophic foundations of quantum mechanics}. Dover, 1944.\\
\\
von Neumann, John: \textit{Mathematische Grundlagen der Quantenmechanik}. Springer, 1932; \textit{Mathematical Foundations of Quantum Mechanics}. Princeton University Press, 2018

}

\end{document}